\documentclass{aa}          
\usepackage{graphicx}
\begin{document}
   
\title{Intra-night Optical Variability of Luminous Radio Quiet QSOs}


\author {A. C. Gupta$^{1,2,3}$, \& U. C. Joshi$^2$}

\offprints{A. C. Gupta}
   
\institute 
{$^1$ Harish-Chandra Research Institute, Chhatnag Road, Jhunsi, Allahabad - 211 019, India \\
$^2$ Astronomy and Astrophysics Division, Physical Research Laboratory, Navrangpura, 
Ahmedabad - 380 009, India \\
$^3$ Tata Institute of Fundamental Research, Homi Bhabha Road, Colaba, Mumbai - 400 005, India 
(present address) \\
email: alok@tifr.res.in, joshi@prl.ernet.in}

\date{Received; Accepted}

\titlerunning{Variability of RQQSOs}
   
\abstract{  

In the present paper we report the detection of intra-night variability in some
of the radio-quiet quasi-stellar objects (RQQSOs) and one lobe dominated
radio-loud quasi-stellar object (LDQ). To study intra-night variability, we
carried out  photometric monitoring of seven RQQSOs and one LDQ in Johnson
V-passband using  1.2 meter optical/IR telescope at Gurushikhar, Mount Abu,
India. Observations were  made in nine nights during the first half of the year
2000; seven RQQSOs: 0748+291,  0945+438, 1017+280, 1029+329, 1101+319,
1225+317, 1252+020 and one LDQ: 1103-006  were observed. RQQSOs 0748+291,
1225+317 and LDQ 1103-006 have shown existence of intra-night variations. In 
the case of 1017+280 (RQQSO) there is indication  of intra-night
variation in one night where as the observations in another night do not show
convincingly the existence of intra-night variability. RQQSOs 0945+438, 1029+329,
1101+319 and 1252+020 have not shown any intra-night variations. We compiled 
intra-night variability data for radio-loud and radio-quiet AGNs from the literature 
for statistical analysis. It is  found that a good fraction of radio-quiet AGNs
show intra-night variations with the maximum amplitude of variation being about
10\%. On the other hand blazars show at  times intra-night flux variability up
to 100\%. In case of radio-loud AGNs (excluding  blazars), maximum amplitude of
intra-night variation lies between the variability amplitude of radio-quiet
AGNs and blazars i.e. the flux variation is close to 50\%. The results indicate
that the energy generation mechanism and the environment around the central
engine in different classes of AGNs may be similar, if not identical. The standard model
for radio-loud AGNs, where shocks are propagating down relativistic jet or models 
based on disturbances in accretion disks can also explain the micro-variability 
in RQQSOs.


}

\maketitle

\section{Introduction}

There is a general consensus on the dichotomy of quasar population: radio-loud 
(R $>$ 10) and radio-quiet (R $<$ 10) (where R is the ratio of radio (6 cm) to 
the optical (440 nm) flux densities) (Kellerman et al. 1989). It is found that 
$\sim$ 10$-$15 \% of the quasars are in the radio-loud category. There is an 
additional distinction between radio-loud and radio-quiet AGNs: radio-loud 
sources occur in elliptical galaxies and radio-quiet are found to reside in 
galaxies dominated by disk. 

Flux variability is a common property of AGNs. Blazars, in particular, show
variation in the complete electromagnetic spectrum on all time scales ranging 
from minutes to years (e.g. Miller et al. 1989; Quinn et al. 1996; Heidt \& Wagner 
1996; Catanese et al. 1997; Lamer \& Wagner 1998; Fan \& Lin 1999; Kataoka et al. 
1999; Peng et al. 2000; Petry et al. 2000; Ghosh et al. 2000; Pursimo et al. 2000; 
Fan et al. 2002; Gupta et al. 2002, 2004; Gupta \& Joshi 2005; Sagar et al. 2004; 
Villata et al. 2004 and references therein). 

Variability time scales can broadly be divided into three classes: (i) flux 
variability in few minutes to few hours is generally known as micro-variability or 
intra-night variability or intra-day variability; (ii) time scales ranging from days 
to weeks can be classified as short term variability; (iii) and the time scale 
ranging from months to years can be called as long term variability.

The first report of micro-variability in AGNs can be found in Mathews \& Sandage (1963), 
who observed a 0.04 mag change in 15 minutes in 3C48. Subsequent to this there are 
several reports on the detection of rapid flux variations in optical region in AGNs 
(Mathews \& Sandage 1963; Oke 1967; Racine 1970; Angione 1971; Bertaud et al. 1973; 
Miller 1980). However, these results were not taken seriously as such small amplitude 
variations might be due to instrumental errors. First convincing case for optical
micro-variability is reported for BL Lacertae using CCD detector by Miller et al. (1989). 
Since then extensive observations using CCD have led to unambiguous confirmation of 
optical micro-variability for a large number of blazars (e.g. Miller et al. 1989; 
Carini et al. 1990; Carini \& Miller 1992; Heidt \& Wagner 1996; Noble et al. 1997; 
Ghosh et al. 2000; Romero et al. 2002; Sagar et al. 2004; Gupta \& Joshi 2005 and 
references therein). All blazars which exhibit micro-variations are found to be 
radio-loud sources and it is believed that relativistic jet dominate their emission 
(Bregmann 1992). Presently, it is accepted that not only the blazars but many other
radio-loud quasars exhibit intra-night variations (Jang \& Miller 1995, 1997; 
Wagner \& Witzel 1995; de Diego et al. 1998; Romero et al. 1999; Stalin et al. 2004, 
2005 and references therein). However, detection of intra-night variations in 
radio-quiet AGNs has been elusive and little is known about their intra-night 
variability. In recent past there have been attempts by several groups around the 
globe to find the intra-night variability in different sub-classes of radio-quiet 
AGNs (Jang \& Miller 1995, 1997; Anupama \& Chokshi 1998; de Diego et al. 1998; 
Romero et al. 1999; Petrucci et al. 1999; Gopal-Krishna et al. 2000, 2003; 
Stalin et al. 2004, 2005). In most of the cases, the reports on micro-variability 
are not overwhelmingly convincing, though in a few RQQSOs intra-night peak-to-peak 
variation of $\sim$ 1\% is reported (Gopal-Krishna et al. 2003; Stalin et al. 2004, 
2005). The results reported by various groups have in fact created some confusion 
on the intra-night variability in radio-quiet AGNs. 

One of the important motivation to study variability in AGNs is to know the
physical scales of the emitting regions. As it is quite difficult to resolve
the nuclei of AGNs with the present day technology, a reasonable way to
investigate the structure and physical conditions near the nucleus is to
study micro-variations of flux and degree of polarization. It is believed
that radio-quiet AGNs either do not have relativistic jet (Antonucci et
al. 1990) or harbor very weak jet (Miller et al. 1993; Kellermann et al.
1994) and hence the effect of jet is expected to be negligible. The presence
of micro-variability in radio-quiet AGNs is, therefore, attributed to
the disturbances on the accretion disk (e.g. hot spots or flaring). For
radio-loud AGNs, both the shocked jet and disturbances on accretion disk may
be responsible for micro-variability. Therefore comparison of
micro-variations between radio-quiet and radio-loud AGNs could constrain
some of the existing models. Micro-variability may be observed as discrete
events or as part of a longer duration variation. The importance of
micro-variability resides in the fact that, if it is intrinsic to the
source, it provides limits on the size of the emitting regions, providing a
powerful tool to investigate both the physical structure of the central
regions of AGNs and the processes responsible for the  production of the
extreme luminosities observed for these objects. The detection of 
micro-variability on a time scale of hours in radio-quiet AGNs is considered
to be a powerful discriminator between accretion disk and relativistic jet
models of these sources.  

In the present study a sample of seven bright RQQSOs and one bright LDQ have
been considered for the study of micro-variations. Observations were carried
out during the period January$-$April, 2000 (nine observing nights) and
the results are reported in this paper. 

The paper is structured as follows: section 2 presents the details about
target selection criterion, in section 3 we report observation and data
reduction techniques, in section 4 results of the present work and
statistical analysis of the previous work are presented and in section 5
conclusions are given.

\section {Target Selection Criterion}

\begin{table*}[t]
\caption[]{Complete log of V band observations of seven radio quiet QSOs and one lobe 
dominated quasars from
1.2 meter Gurushikhar Telescope at Mount Abu, India}
\begin{center}
\begin{tabular}{cccccccccc}
\hline \hline \\
IAU Name$^{*}$ & Other Name  & $\alpha_{2000.0}$ & $\delta_{2000.0}$ & z & V & M$_{V}$ & Date of & Data & Duration \\
               &               &                &            &      &       &      & Observations & Points & (hours) \\
               &               &                &            &      &       &      & dd.mm.yyyy & & \\\hline
0748$+$291  & QJ 0751$+$2919   & 07 51 12.3     & $+$29 19 38      & 0.912 & 16.14 & $-$27.9 & 13. 01. 2000 & 42 & 8.0 \\
0945$+$438  & US 995           & 09 48 59.4     & $+$43 35 18      & 0.226 & 16.28 & $-$24.5 & 26. 02. 2000 & 26 & 5.5 \\
1017$+$280  & Ton 34           & 10 19 56.6     & $+$27 44 02      & 1.918 & 15.69 & $-$29.8 & 14. 01. 2000 & 39 & 6.5 \\
1017$+$280  & Ton 34           & 10 19 56.6     & $+$27 44 02      & 1.918 & 15.69 & $-$29.8 & 27. 02. 2000 & 55 & 6.1 \\
1029$+$329  & CSO 50           & 10 32 06.0     & $+$32 40 21      & 0.560 & 16.00 & $-$26.7 & 05. 04. 2000 & 61 & 6.0 \\
1101$+$319  & Ton 52           & 11 04 07.0     & $+$31 41 11      & 0.440 & 17.30 & $-$24.9 & 04. 04. 2000 & 47 & 6.4 \\
1103$-$006  & PKS 1103$-$006   & 11 06 31.8     & $-$00 52 53      & 0.426 & 16.46 & $-$25.7 & 06. 04. 2000 & 48 & 5.7 \\
1225$+$317  & b2 1225+317      & 12 28 24.8     & $+$31 28 38      & 2.219 & 15.87 & $-$30.0 & 07. 04. 2000 & 54 & 6.2 \\
1252$+$020  & q 1252+020       & 12 55 19.7     & $+$01 44 12      & 0.345 & 17.30 & $-$24.4 & 09. 03. 2000 & 25 & 3.7 \\
\hline 
\end{tabular}
\end{center} 
$^{*}$ based on coordinates defined for 1950.0 epoch
\end{table*}

The radio quiet QSOs and LDQ for the present study were selected from the lists
of V$\acute e$ron-Cetty \& V$\acute e$ron (2001). Detailed information of
the seven RQQSOs and one LDQ and their dates of observations are listed in 
Table 1. Hubble constant H$_{0}$ = 50 km s$^{-1}$ Mpc$^{-1}$ and q$_{0} =$ 0.5
are assumed for determining M$_{V}$. 

Simultaneous observation of the target source and a few comparison stars
and the sky background allow to remove variations which may be due to
fluctuations in either atmospheric transparency or extinction. Therefore,
RQQSOs and LDQ were selected for observation in such a way as 
to have at least two comparison stars in the field of view of the 
camera with brightness comparable to the target source.

Carini et al. (1991) investigated whether a conspicuous galaxy component produce 
variations due to fluctuations in atmospheric seeing or transparency which are 
not intrinsic to the source. They showed that even for sources with significant 
underlying galaxy components, any spurious variations introduced by fluctuations 
in atmospheric seeing or transparency are typically smaller than the observational 
uncertainties. To further reduce this effect, we have selected sources which are 
optically bright (brighter than M$_{V} < -$24.4 mag) so that the fluctuations due 
to the underlying galaxy are minimal. The modest optical luminosities 
(M$_{V} > -$24.4 mag) lie close to the critical value below which the sources 
become like those of Seyfert galaxies (Miller et al. 1990). At these lower 
levels of AGN to galactic light ratios, 
false indications of variability produced 
by seeing variations that include different amounts of host galactic light within 
the photometric aperture, become very probable (Cellone et al. 2000).

In our sample all the sources are brighter than M$_{V} \leq -$24.4 (vide. Table 1) 
thus minimizing the seeing effects. The host galaxy is expected to contribute less 
than 10\% to the total flux of the luminous RQQSOs or the LDQ. The host galaxy is
also expected to be encompassed within the aperture used for photometry.

\section {Observations and Data Reductions}

CCD photometric monitoring of seven RQQSOs and one LDQ were carried out in Johnson 
V-passband using a thinned back illuminated Tektronix 1K $\times$ 1K CCD detector 
at f/13 Cassegrain focus of 1.2 meter Gurushikhar Telescope, Mount Abu, India. 
To improve signal to noise ratio (S/N), on CCD chip binning (2 $\times$ 2) was done 
while reading out the array. One super pixel projected on the sky corresponds to 
0.634 arcsec in both the dimensions. Entire CCD chip covers $\sim$ 5.4 $\times$ 5.4 
arcmin$^{2}$ of the sky. Read out noise and gain of the CCD detector were 4 electrons 
and 10 electrons/ADU respectively. Throughout the observing run, typical average seeing 
(FWHM of stellar image) was $\sim$ 1.5 arcsec ranging between 1.2 to 1.8 arcsec. Several 
bias frames were taken intermittently in each observing night and sky flats in V-passband 
were taken during the twilight hours. 
 
Image processing or initial processing (bias subtraction, flat-fielding and cosmic 
rays removal), photometric reduction or processing (getting instrumental magnitudes 
of stars and target RQQSOs or LDQ in the image frames) were done at Physical Research 
Laboratory, Ahmedabad, India and at Harish-Chandra Research Institute, Allahabad, 
India using IBM $-$ 6000 RISC workstations and Pentium III computers.

\begin{table}[h]
\caption[]{Comparison star locations (relative to QSOs)}
\begin{tabular}{ccccccc}
\hline \hline \\
QSO    &  Star 1 &  &  Star 2 & &  Star 3 &   \\
       & $\Delta$r$\arcsec$ & PA$^{0}$ & $\Delta$r$\arcsec$  & PA$^{0}$ & $\Delta$r$\arcsec$ & PA$^{0}$ \\\hline
0748$+$291         & 128 & 352 & 93 & 332 &&\\
0945$+$438         & 15  & 340 & 95 & 220 &&\\
1017$+$280         & 208 & 38  & 107 & 311 & 135 & 78 \\
1029$+$329         & 162 & 263 & 246 & 153 & &\\
1101$+$319         & 167 & 169 & 201 & 20 &&\\
1103$-$006         & 283 & 236 & 37  & 16 &&\\
1225$+$317         & 142 & 242 & 196 & 334 &&\\
1252$+$020         & 301 & 229 & 223 & 161 &&\\
\hline 
\end{tabular}
\end{table}

Standard routines in the IRAF package were used for the initial processing of the images.
Median bias frames and flat-field images were constructed for each night which were used
for bias and flat field correction. Instrumental magnitude of RQQSOs, LDQ and comparison 
stars in the RQQSOs and LDQ fields were determined by using DAOPHOT II software 
(Stetson 1987, 1992) and concentric aperture photometric technique. Aperture photometry 
was done with several concentric aperture radii: 3.5, 5.0, 7.0, 9.5 and 12.0 pixels. 
Though the data reduced with different aperture radii are found to be in good agreement, 
aperture radius of 7 pixels gives the best S/N and therefore the photometric magnitudes 
reported here are based on that aperture radius. Stars in different frames of same source 
were matched by using DAOMATCH routine in DAOPHOT II package. The differential magnitudes 
were calculated for pairs of stars on the frame. Two comparison stars (non variable 
during our observing run) were used to generate the differential light curves of RQQSOs 
and LDQ. The positions of the comparison stars in RQQSOs and LDQ fields are listed in 
Table 2.

Simultaneous observations of the variable source and a few comparison stars and the sky 
background allow to remove variations which may be due to fluctuations in either the 
atmospheric transparency or the extinction. One may also find the gradual variation in 
differential magnitude as function of zenith distance if the colour of the target
source and the comparison stars differ very much. The data reported here were obtained 
during the good photometric quality nights and also it was insured that the zenith angle 
for observations do not exceed 60\degr. This reduces the colour dependence of differential 
magnitude on zenith angle. Carini et al. (1992) examined the plots of differential 
magnitude between comparison stars of different colours versus airmass, and found that 
over a large range of airmass, neither the overall photometric accuracy was affected 
significantly by the large colour difference in the sets of comparison stars, nor did 
it introduce systematic variations not intrinsic to the source. Also a closer look at 
the data do not indicate any signature of colour dependence with zenith distance.

\section{Results}

\subsection {Differential Light Curves (DLCs)}

DLCs of seven RQQSOs and one LDQ which were observed during nine nights in the V passband 
are plotted in Figure 1$-$9. In the following, we discuss the criteria to test the existence 
of variability. We also mention here that the individual bias corrected and flat fielded 
images of targets were examined carefully to see if there was any background variation 
after initial processing of the images and image frames which showed gradual variation 
(say more than 2\% end to end) in CCD response were rejected. Lastly it is most important 
that the DLCs of any target should show good correlation so that there be no doubt on the 
variability of the source.

\subsection {Variability Detection Criterion}

We have followed the method outlined by Howell et al. (1988) to detect objectively
the presence or absence of variability at a particular confidence level (say
$3\sigma$ or above).  In the present study, for analysis of each source, two
comparison stars (s1 and s2) were used  and DLCs were generated. We estimate rms
error by fitting a straight line to the DLC of comparison stars (comparison star -
check star) (s1-s2) using the least square fit and estimate  the deviation of the
data points from the fitted straight line. The mean value of the standard 
deviation has been used as the measure of the observational error. The formal
error for each  data point are substantially smaller than the standard deviation
($\sigma$) of the  comparison - check star data and therefore is much more
generous estimate of the true  observational error than the formal errors (photon
noise) detected by photometry software  DAOPHOT II. In general the luminosity of
the comparison stars is different than the target  source. The value of the
standard deviation estimated as above was scaled to the $\sigma^2_{v-s}$  by using
the equation (4) of Howell et al. (1988). The value of $\Gamma^{2}$ was calculated
using  equation (13) in Howell et al. (1988) which is used to scale the
$\sigma^{2}_{s1-s2}$ to get  $\sigma^{2}_{v-s}$.  The scaled value
of $\sigma^2_{v-s}$ was then used to assess the confidence level of variability.

\subsection {Notes on different sources:} 

\noindent
{\bf 0748+294 (QJ 0751+2991)} \\
\\
This RQQSO, reported as the brightest new QSO in the first bright QSO survey
(Gregg et al. 1996), was monitored on the night of February 14, 1999 in the
optical R band by Gopal-Krishna et al. (2000) to search for micro-variability. 
In the seven hours of continuous monitoring of the source, they have reported 
spikes (excursion of just one point) at two occasions and suggested the possible 
existence of micro-variability in the source. Observations of this source in 
six nights with average monitoring ($\sim$ 6.5 hours per night) in R passband 
is again reported by Stalin et al. (2004) and they did not find micro-variability 
in this source but spike of $\sim$ 2\% brightness excursion was seen on one occasion. 
However, Stalin et al. (2004) monitored this source for more than three years 
(December 14, 1998 to December 25, 2001) and on the basis of the data they have 
reported the existence of long term variation in the source.  

\vspace*{-0.5in}
\begin{figure}[h]
\includegraphics[bb=18 144 592 718,width=4.3in,height=3.4in,clip]{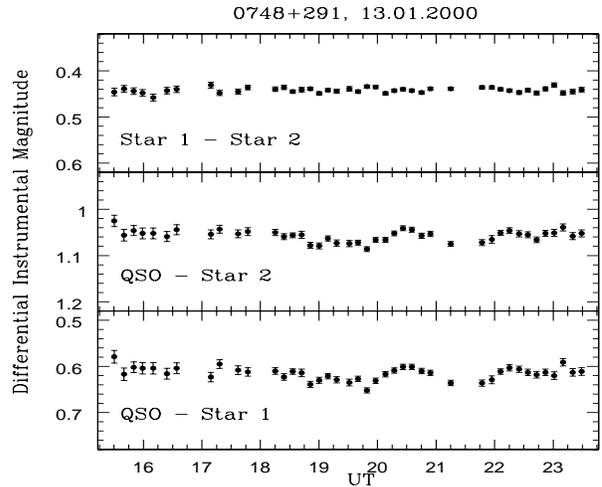}
\caption[]{ The $V$ band light curve of 0748+294 on the night of January 13, 2000.} 
\label{fig}
\end{figure}

We observed this source continuously for about eight hours (UT 15.669 to 23.336
hr) during the night of January 13, 2000; data sampling being about 5 points per
hour and integration time for each frame being 500 sec. DLCs were obtained with
respect to two comparison stars, both the comparison stars being brighter than
the source. Both the DLCs (QSO - star1 and QSO - star2) for the source 0748+291
are plotted in Fig. 1 which clearly indicate brightness variation of about 4\%
during the time UT 18.553 - 22.569 hr. As is clear from Fig. 1, the DLC for the
comparison stars (star1 - star2) is quite steady whereas the DLCs for the
source show appreciable variation during the period. To quantify the degree of
variation, we estimated the variance from the DLC for the comparison stars
assuming that the stars remained steady during the observing run. A straight
line was fitted to the data ($\Delta$ mag against UT) and the deviations are
estimated from the mean line. The rms noise (standard deviation) is estimated
at 0.004 mag. This is scaled to the DLC of the (source - comparison)  using the
method stated above in section 4.2 and the rms noise $\sigma$ is estimated at
0.005 mag. During the period UT 18.553 - 22.569 hr, rms variation of
qso-comparisons are respectively 0.014 and 0.013 for qso - s1 and qso - s2. Thus
the variance for DLCs of qso is more than 6 $\sigma$.  
This indicates detection of genuine micro variation in the source. 
If we consider peak-to-peak variation ($\approx$ 0.05 mag) in the DLC of
qso - star1, the variation is at a level of 10 $\sigma$, which further support
the existence of micro-variability in the source. \\

\noindent
{\bf 0945+438 (US 995)} \\
\\
Huang et al. (1990) studied the variability of this source using the data taken from 
Palomar plates for the period 1978 to 1981 ($\sim$ 3.5 year) and did not find any 
variability in the source. This source was also observed in near-infrared pass-bands
(JHK$^{'}$) during the period February 10, 1996 and December 27, 1997 by Enya et al.
(2002) and they have reported the existence of long term variation which they were 
interested in. Search for micro-variability in this source in optical bands was done 
by different groups e.g. de Diego et al. (1998) in two nights, Stalin et al. (2004) 
in three nights, Stalin et al. (2005) in one night. de Diego et al. (1998) have 
reported the existence of micro-variability in the source but on other hand Stalin 
et al. (2004, 2005) did not find any evidence of micro-variability. However, the 
existence of long term variation in the source has been reported by Stalin et al.
(2004), the source has dimmed about 0.07 mag during February 26, 2000 and January 23, 
2001. 
 
\vspace*{-0.5in}
\begin{figure}[h]
\includegraphics[bb=18 144 592 718,width=4.3in,height=3.4in,clip]{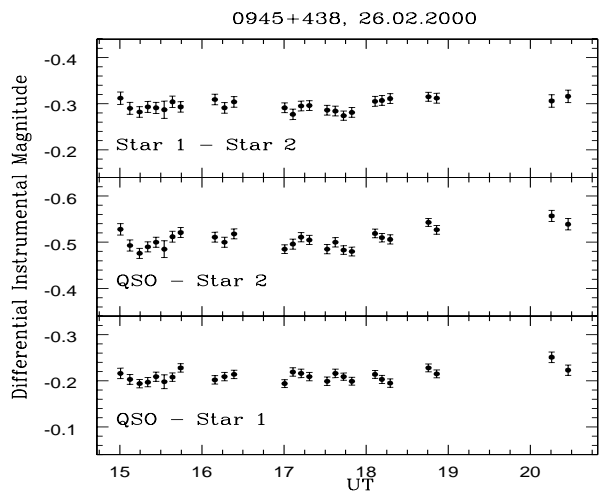}
\caption[]{The $V$ band light curve of 0945+438 on the night of February 26, 2000.} 
\label{fig}
\end{figure}

We observed this source for more than 5 hours during (UT 15.125 - 20.458 hour) on
the night of  February 26, 2000. Integration time for each frame was 300 seconds.
The DLCs are shown in Fig. 2. There is a break in observations during 18.9 - 20.2
hours UT due to some problem in the control system of the  telescope. DLC of
comparison stars gives $\sigma$ at 0.011 and scaled value is 0.010.  Standard
variation of qso-comparisons are respectively 0.011 and 0.018 for qso - s1 and 
qso - s2, and hence the mean variance is less than 2 $\sigma$. The data
appear a bit noisy.  Hence, the source has not shown any significant variation 
during the night. \\ 

\noindent 
{\bf 1017+280 (Ton34)}  \\ 

Stalin et al. (2004) have monitored this source to look for the existence  of
micro-variability and their observations in three nights in R-band did not show
any clear evidence of micro-variability.

\vspace*{-0.5in}
\begin{figure}[h]
\includegraphics[bb=18 144 592 718,width=4.3in,height=3.4in,clip]{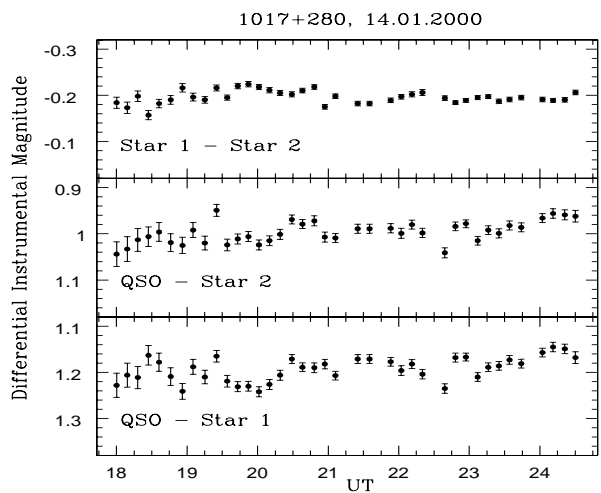}
\caption[]{ The $V$ band light curve of 1017+280 on the night of January 14, 2000.}
\label{fig}
\end{figure}

\vspace*{-0.5in}
\begin{figure}[h]
\includegraphics[bb=18 144 592 718,width=4.3in,height=3.4in,clip]{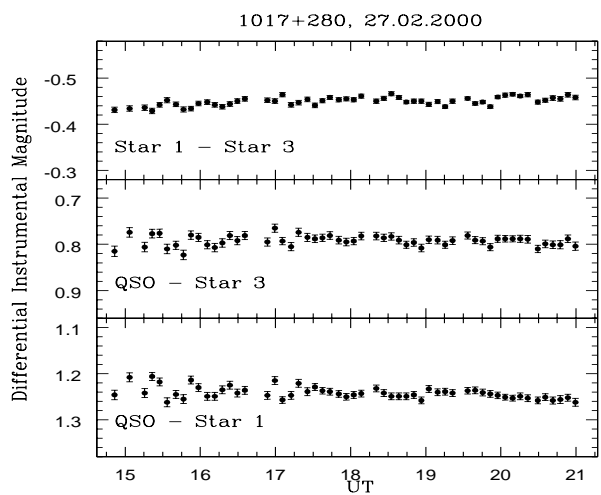}
\caption[]{ The $V$ band light curve of 1017+280 on the night of February 27, 2000.}
\label{fig}
\end{figure}

We observed this source on two nights, January 14 and February 27, 2000, 
for continuous 6.5 and 6.0 hours respectively. On January 14, 2000 integration
time for each image frame was 400 seconds whereas it was 300 seconds for the
observing run on February 27, 2000. The DLCs are plotted in Fig. 3 and Fig. 4. 
During the night of January 14, DLC of the comparisons is not stable during the
period UT 18.003 - 21.586 hr,  but later on it is relatively stable. Hence, we 
consider only the data during the period of UT 21.886 - 24.503 hr for further
discussion.  The standard
deviation of comparison stars is estimated at 0.006 mag, scaled value of
$\sigma$ for DLCs of the source is 0.008. The standard deviation for DLCs of
qso - s1 and qso - s2 are respectively 0.019 and 0.018. The mean variance for
the DLCs of the source is more than  5 $\sigma$. This indicates that the source
is showing variability during the specific period of the night. 

In the night of February 27, 2000, DLC of comparisons is quite stable the
$\sigma$ value is estimated at $\sim$ 0.007,
and scaled value for source DLC is  0.010. The standard variations for DLCs of
qso are 0.011 and 0.011. Hence the  variance is less than 2 $\sigma$.
Therefore, no variation is detected during the night.  \\

\noindent
{\bf 1029+329 (CSO 50)} \\
\\
This source was observed earlier by other groups in six nights (Gopal-Krishna et al. 2003; 
Stalin et al. 2004, 2005) and the existence of micro-variability has been reported in 
two nights out of the six nights. However, no significant long term variation in the span
of two years are reported.

\vspace*{-0.5in}
\begin{figure}[h]
\includegraphics[bb=18 144 592 718,width=4.3in,height=3.4in,clip]{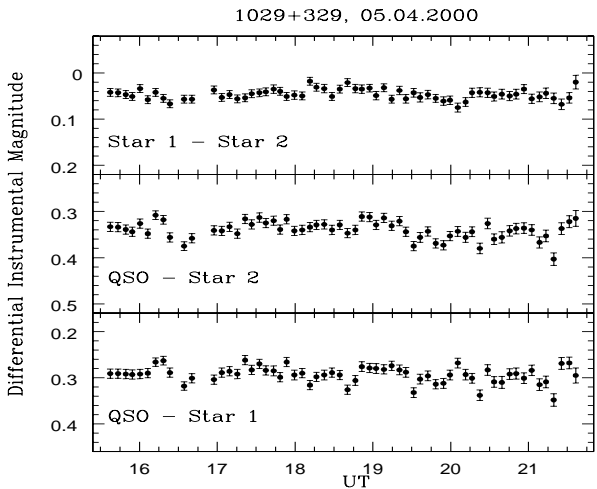}
\caption[]{ The $V$ band light curve of 1029+329 on the night of April 05, 2000.}
\label{fig}
\end{figure}

We observed this source continuously for about 6 hours on the night of April 05,
2000.  The integration time for each image frame was 300 sec. The DLCs are plotted
in Fig. 5.  The $\sigma$ value for the DLC of the comparisons is estimated $\sim$
0.011 which gets scaled to the same value as $\Gamma$ is close to 1; $\sigma$
values for qso - s1 and qso - s2 are respectively 0.017 and 0.018.
As the observed variance for the source is close to 2 $\sigma$. Hence, no 
micro variability is detected. \\

\noindent
{\bf 1101+319 (Ton 52)}\\
\\
Stalin et al. (2004) observed this source in four nights for searching 
micro-variability and have reported the existence of micro-variability in one 
night, long term variation is also seen in the source. 

\vspace*{-0.5in}
\begin{figure}[h]
\includegraphics[bb=18 144 592 718,width=4.3in,height=3.4in,clip]{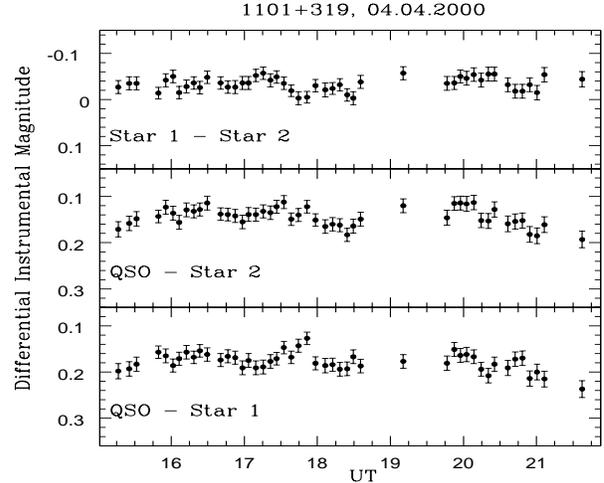}
\caption[]{ The $V$ band light curve of 1101+319 on the night of April 04, 2000.}
\label{fig}
\end{figure}

We also observed this source for about 6 hours on the night of April 04, 2000.
Integration  time for each image frame was 300 sec. The DLCs are plotted in
Fig. 6. From the plot it appears that the quality of data is relatively
poor and data are a bit noisy. The scaled  $\sigma$ value of the comparison
stars is estimated at 0.013. The $\sigma$ values  for DLCs of qso are
respectively 0.019 and 0.020 for qso - s1 and qso - s2. The variance for the
DLCs of qso is close to 2 $\sigma$, indicating the non existence of 
micro-variability in the source.  \\
 
\noindent 
{\bf 1103-006 (PKS 1103-006)} \\

This is the only lobe dominated quasar in our sample. The source was observed
by Enya et al. (2002) in near infrared pass-bands JHK$^{'}$ for searching long
term variability. Observations  were made  during the period February 12, 1996
and January 05, 1998. The source was found  showing variation on long term.
This source was also observed by Stalin et al. (2004) during the period March
17, 1999 to March 22, 2002 (six nights) in optical region and have reported 
clear evidence of micro-variability in one night and also reported significant
variation in long term.

We observed this source for about 5.5 hours on the night of April 06, 2000.
Integration  time for each image frame was 300 sec. The DLCs are plotted in
Figure 7. Scaled value of $\sigma$ from the DLC of comparison stars is
estimated  $\sim$ 0.015. From the DLCs of the  source, the standard deviations
are 0.026 and 0.020 respectively for DLCs  qso - s1 and qso -  s2. Thus the
variance for the source is about 3 time the variance of the standards,
indicating the possible existence of intra-night variation.

\vspace*{-0.5in}
\begin{figure}[h]
\includegraphics[bb=18 144 592 718,width=4.3in,height=3.4in,clip]{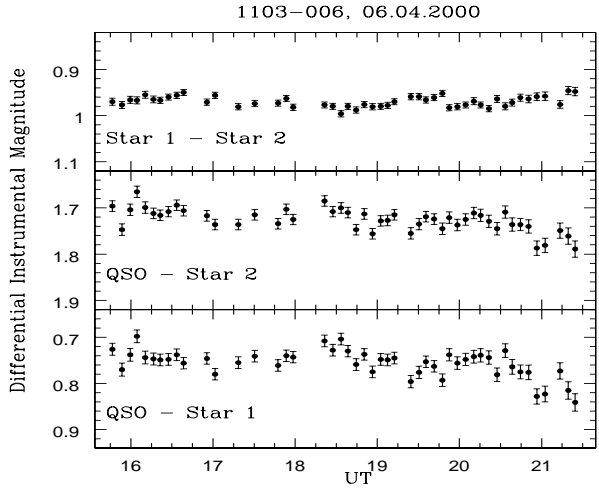}
\caption[]{ The $V$ band light curve of 1103-006 on the night of April 06, 2000.}
\label{fig}
\end{figure}

\noindent
{\bf 1225+317 (b2 1225+317)} \\

\vspace*{-0.5in}
\begin{figure}[h]
\includegraphics[bb=18 144 592 718,width=4.3in,height=3.4in,clip]{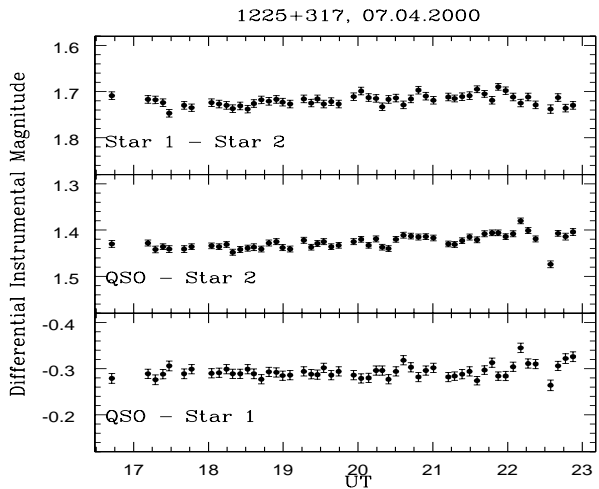}
\caption[]{ The $V$ band light curve of 1225+317 on the night of April 07, 2000.}
\label{fig}
\end{figure}

There has been no systematic attempt earlier to study this source for
micro-variability. We monitored this source for about six hours (UT 15.669 to
23.480 hr) during the night of April 07, 2000. Integration time for each image
frame was 300 sec. DLCs are plotted in Fig. 8. The $\sigma$ value based on DLC of
comparison stars is estimated at 0.011 which scales to 0.008 for source DLCs. 
DLCs for the source show standard deviation 0.016 and 0.012 for qso - s1  and qso
- s2 respectively. The variance for the DLC of qso is thus close to 3 times the
$\sigma$ value, indicating the possible existence of micro variation. During  the
period UT 22.175 - 22.575 hr  DLCs of qso - s1 and qso - s2 show brightness
variation 0.081 mag and 0.094 mag respectively which is close to 10 $\sigma$
level. Confirmation of such events require further monitoring of the
source with larger S/N.  \\

\noindent
{\bf 1252+020 (q 1252+020)}\\

\noindent 
This source was observed for five nights during March 22, 1999 to March 18,
2002 by Stalin et al. (2004). Source has shown micro-variability in two
nights and significant long term variability is also reported.

This source was observed by us for about 3.5 hours in the night of March 09, 2000.
The DLCs are plotted in Fig. 9.  The standard deviation based on DLC of comparison stars 
scaled to source DLC is estimated at 0.014. 
Standard deviations for the DLCs of qso are respectively 0.014 and 0.014 for
qso-s1 and qso-s2,  indicating the absence of micro variation in the source. 

\vspace*{-0.5in}
\begin{figure}[h]
\includegraphics[bb=18 144 592 718,width=4.3in,height=3.4in,clip]{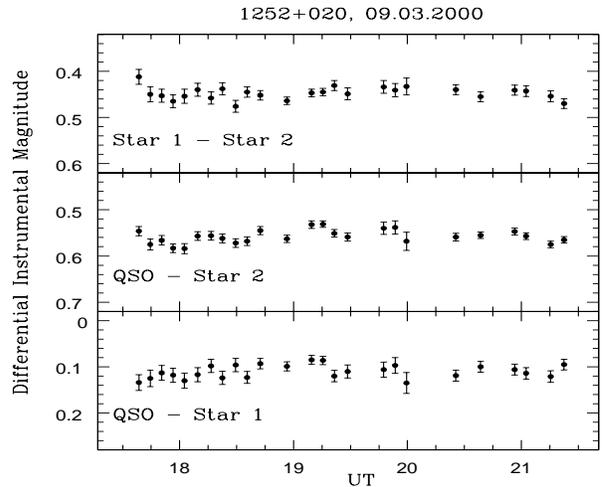}
\caption[]{ The $V$ band light curve of 1252+020 on the night of March 09, 2000.}
\label{fig}
\end{figure}

\subsection {Statistical Analysis of Intra-night Optical Variability of Different 
Classes of AGNs}

\subsubsection {Radio-Quiet AGNs}

\begin{table*}[t]
\caption[]{Log of radio-quiet AGNs which were monitored by various researchers 
looking for intra-night variability in the optical bands: column 1 - number of light 
curves (LCs) available; column 2, 3, 4 - number of LCs available for the duration 
indicated at the top of the respective columns. The numbers in the brackets indicate 
the number of events when the source has been reported: non variable, possible variable, 
and variable respectively; column 5 - source of the data (reference).}
\begin{center}
\begin{tabular}{ccccc}
\hline \hline \\
No. of LCs \hspace*{0.3in} & \hspace*{0.3in}      & \hspace*{0.3in} Radio Quiet AGNs \hspace*{0.3in} &       & \hspace*{0.3in} Ref. \\\hline
        &  duration $\leq$ 3h & 3h $<$ duration $\geq$ 6h &  duration $>$ 6h &
\hspace*{0.3in} \\\hline
27      & 0(0,0,0)   & 26(22,2,2)      & 1(1,0,0)   & 1 \\
2       & 2(0,0,2)   & 0(0,0,0)        & 0(0,0,0)   & 2 \\
10      & 0(0,0,0)   & 3(3,0,0)        & 7(6,1,0)   & 3 \\
23      & 8(8,0,0)   & 15(15,0,0)      & 0(0,0,0)   & 4 \\
55      & 23(15,7,1) & 19(11,8,0)      & 13(7,6,0)  & 5 \\
29      & 0(0,0,0)   & 13(11,0,2)      & 16(13,0,3) & 6, 7 \\
20      & 6(4,0,2)   & 8(5,0,3)        & 6(5,0,1)   & 8 \\
8       & 0(0,0,0)   & 2(2,0,0)        & 6(3,1,2)   & 9 \\\hline
174     & 39(27,7,5) & 86(69,10,7)     & 49(36,8,6) & {\bf Total} \\\hline
\end{tabular}
\end{center} 
{\bf Ref.} (1) Jang \& Miller (1997); (2) Anupama \& Chokshi (1998); (3) Romero et al. (1999); 
(4) Petrucci et al. (1999); (5) Gopal-Krishna et al. (2000); (6) Gopal-Krishna et al. (2003);
(7) Stalin et al. (2004); (8) Stalin et al. (2005); (9) present study 
\end{table*}

During the last decade several groups have done extensive search for finding 
micro-variability in different subclasses of radio-quiet AGNs. The results based 
on the study made by various researchers is described briefly in the following. 

A sample of 19 radio-quiet AGNs was studied by Jang \& Miller (1995, 1997) for searching 
micro-variability. They presented DLCs for RQQSOs Ton 951 and Ton 1057 which show up to 
$\sim 8\%$ variation on time scale of an hour or so. However, both these sources are of 
the modest luminosity (M$_B > -24.3$) and at these lower level of luminosity, false 
indications of variability, due to varying seeing, cannot be ruled out (Cellone et al. 
2000). Jang \& Miller (1995, 1997) have reported 16\% (3/19) sources showing
micro-variability. Their statistical analysis show the existence of variability at 
confidence level of 99\%.

\begin{figure}[h]
\hspace*{-0.2in}
\includegraphics[bb=18 144 592 718,width=4.0in,height=4.0in,clip]{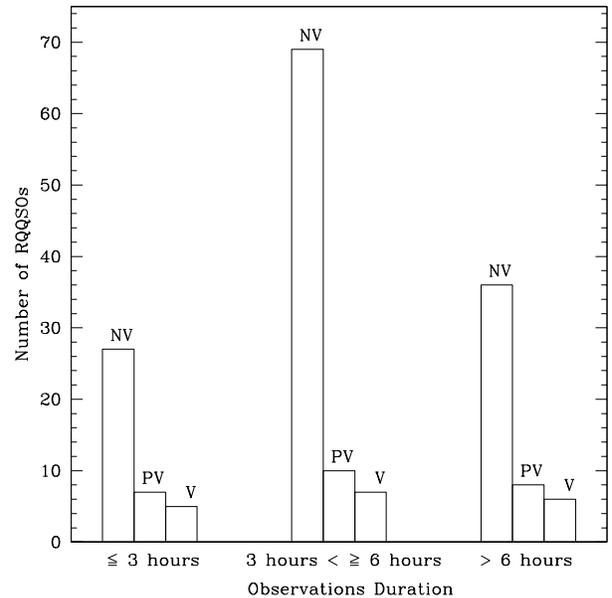}
\caption[]{Histogram of observing runs on radio-quiet AGNs. NV, PV and V denote the 
numbers of events the sources were detected: non variable, possible variable and variable 
respectively.}
\label{fig}
\end{figure}

de Diego et al. (1998) monitored a sample of 34 sources, equally distributed between 
radio-quiet and core dominated radio-loud QSOs. They observed pair of objects from both 
the categories having reasonably matched redshifts and luminosities. Based on the variability  
behavior, they claim that there is no difference in radio-loud and radio-quiet QSOs. However,
their data is rather scanty as each source was monitored only a few times per night and 
there was no attempt to systematically monitor sources continuously for a few hours 
(say for about three hours or more). So, the data lacks the continuity of a lengthy data 
trains. 
We have not considered these results in the statistical analysis in the present paper.

Rabbette et al. (1998) failed to detect intra-night variability in a sample of 
23 high-luminosity RQQSOs. Their detection threshold was $\sim$ 0.1 mag. Their data 
lacked the continuity of a lengthy data trains 
and not used for statistical analysis in the present paper.  

Search for rapid optical variability in two broad-absorption line QSOs (BALQSOs) was 
conducted by Anupama \& Chokshi (1998) and they have reported detection of significant 
variation in both the QSOs. 

Romero et al. (1999) observed a sample of 23 southern AGNs in which eight were RQQSOs and 
rest were belonging to the different subclasses of radio-loud AGNs. In their analysis, they  
used the scatter in the weighted average of six comparison stars for estimating photometric 
errors. None of their eight RQQSOs showed indications strong enough to support the existence 
of of intra-night variability.

A micro-variability study of 22 Seyfert 1 galaxies (relatively weak, radio-quiet AGNs) have 
been done by Petrucci et al. (1999). Their error estimation method is different from Romero 
et al. (1999). They took the weighted average of three or more comparison stars to define a 
virtual standard star and used structure function analysis to look for micro-variability in 
these sources. However, they did not find micro-variability in any source in their sample.

Gopal-Krishna et al. (2000) reported the results on micro-variability in a 
sample of 16 RQQSOs. They found 31\% (5/16) to be probable or very probable micro-variable, 
31\% (5/16) RQQSOs showing spikes in their DLCs and the rest 38\% (6/16) sources being 
non variable.

Recently 49 intra-night variability light curves were presented for 19 RQQSOs by Gopal-Krishna
et al. (2003) and Stalin et al. (2004, 2005). They found peak-to-peak micro-variation of
$\sim$ 1\% in 11 light curves of 8 RQQSOs. 11 RQQSOs have not shown any intra-night variations. 

To study the occurrence of micro-variability in radio-quite AGNs and their
statistical behavior, we compiled the data on variability of different
subclasses of radio-quiet AGNs from the literature, thus enlarging the data base.
The statistics is expected to be more robust. The data are listed in Table 3
and the statistics in the form of histogram is plotted in Fig. 10. We find
nearly $\sim$ 10 \% radio-quiet AGNs show intra-night variations.

\subsubsection {Radio-Loud AGNs (Non Blazars)}

\begin{table*}[t]
\caption[]{Log of radio-loud AGNs (excluding blazars) which were monitored by 
various researchers looking for intra-night variability in the optical bands. Details 
on the columns are as given in Table 3.}
\begin{center}
\begin{tabular}{ccccc}
\hline \hline \\
No. of LCs \hspace*{0.3in} & \hspace*{0.3in}      & \hspace*{0.3in} Radio Loud AGNs \hspace*{0.3in} &       & \hspace*{0.3in} Ref. \\
        &        & (excluding Blazars) &     &       \\\hline
        & duration $\leq$ 3h &  3h $<$ duration $\geq$ 6h &  duration $>$ 6h & \\\hline
19      & 1(0,0,1)   & 17(5,1,11)      & 1(1,0,0)   & 1, 2 \\
7       & 0(0,0,0)   &  7(4,0,3)       & 0(0,0,0)   & 3 \\
33      & 2(1,0,1)   & 10(6,0,4)       & 21(14,0,7) & 4 \\
15      & 0(0,0,0)   &  4(3,0,1)       & 11(8,1,2)  & 5 \\
33      & 1(1,0,0)   & 14(12,2,0)      & 18(13,0,5) & 6 \\
8       & 0(0,0,0)   &  5(2,0,3)       & 3(0,0,3)   & 7 \\\hline
115     & 4(2,0,2)   & 57(32,3,22)     & 54(36,1,17) & {\bf Total} \\\hline
\end{tabular}
\end{center}
(1) Jang \& Miller (1995); (2) Jang \& Miller (1997); 
(3) Romero et al. (1999); (4) Romero et al. (2002); (5) Sagar et al. (2004); (6) Stalin et al. (2004); 
(7) Stalin et al. (2005)
\end{table*}

Study of micro-variability in optical wavebands of the radio-loud AGNs 
excluding blazars was carried out by several groups. First systematic search
for optical micro-variations in radio-loud QSOs was carried out by Jang \&
Miller (1995, 1997). They monitored 11 radio-loud  QSOs in 20 nights and found
10 sources showing variation in the flux  at least in one night.

Romero et al. (1999) monitored a sample of 5 radio-loud QSOs in 7 nights.
and found 3 radio-loud QSOs showing flux variation of $\sim$ 2.2 to 8\% within 
a single night. The other 2 radio-loud QSOs have not shown any significant 
variations during the observing run of 4 nights. 

\begin{figure}[h]
\hspace*{-0.2in}
\includegraphics[bb=18 144 592 718,width=4.0in,height=4.0in,clip]{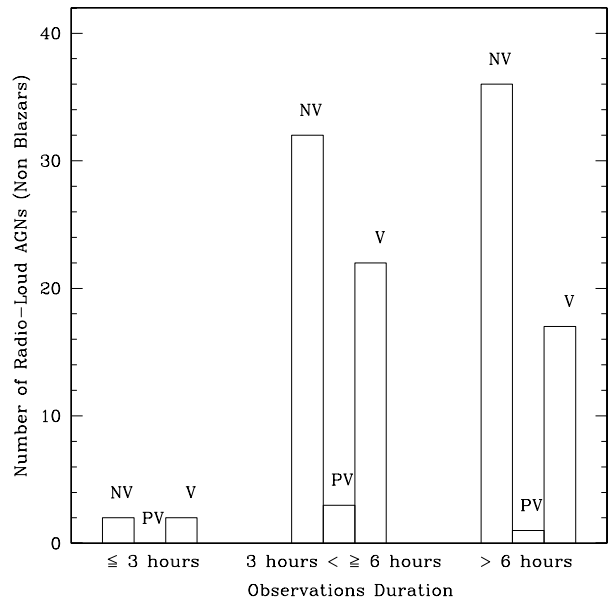}
\caption[]{Histogram of observing runs on radio-loud AGNs (non blazars). NV, PV and V denote 
the numbers of events the sources were detected: non variable, possible variable 
and variable respectively.}
\label{fig}
\end{figure}

Romero et al. (2002) have reported observations of 16 EGRET radio-loud quasars
(non blazars) in 33 nights during the period 1997 to 2000. Intra-night variations 
were reported for 12 nights in 7 radio-loud quasars. 9 radio-loud quasars have 
not shown any intra-night variations.

In a recent paper, Sagar et al. (2004) have reported observations  of 5 core
dominated QSOs (CDQs) in 15 nights in R-band. They found one source showing
variation in one night and another source showed variation in all the three
nights when observations were made. Rest three sources have not displayed any 
intra-night variations in the observations of nine nights.

In another set of recent papers, Stalin et al. (2004, 2005) have reported
optical  monitoring of 5 radio-loud AGNs in 40 nights and have reported
intra-night variations only in 11 nights and one source has shown a possible
intra-night variation on one night.

To investigate the general statistical behavior of micro-variability
of radio-loud AGNs (excluding 
blazars), we compiled the data on variability (based on various monitoring
program to  study micro-variability) from literature which is listed in Table 4.
The data are plotted in the form of histogram in Fig. 11. We find that
nearly $\sim$ 35-40 \% radio-loud  AGNs (non-blazars) show intra-night
variations.

\subsubsection {Radio-Loud AGNs (Blazars)}

\begin{table*}[t]
\caption[]{Log of radio-loud AGNs (blazars) which were monitored by various researchers 
looking for intra-night variability in the optical bands. Details on the columns are as 
given in Table 3.}
\begin{center}
\begin{tabular}{ccccc}
\hline \hline \\
No. of LCs \hspace*{0.3in} & \hspace*{0.3in}      & \hspace*{0.3in} Radio Loud AGNs \hspace*{0.3in} &       & \hspace*{0.3in} Ref. \\
        &        & (Blazars) &     &       \\\hline
        & duration $\leq$ 3h &  3h $<$ duration $\geq$ 6h &  duration $>$ 6h & \\\hline
2       & 0(0,0,0)    & 2(0,0,2)        & 0(0,0,0)   & 1 \\
4       & 0(0,0,0)    & 4(0,0,4)        & 0(0,0,0)   & 2 \\
9       & 8(6,0,2)    & 1(0,0,1)        & 0(0,0,0)   & 3 \\
32      & 13(3,0,10)  & 17(6,0,11)      & 2(1,0,1)   & 4 \\
4       & 0(0,0,0)    & 0(0,0,0)        & 4(0,0,4)   & 5 \\
9       & 5(0,0,5)    & 3(1,1,1)        & 1(1,0,0)   & 6 \\
24      & 2(1,0,1)    & 6(2,0,4)        & 16(3,0,13) & 7 \\
25      & 0(0,0,0)    & 11(7,0,4)       & 14(1,0,13) & 8 \\
4       & 0(0,0,0)    & 2(0,0,2)        & 2(1,0,1)   & 9 \\\hline
113     & 28(10,0,18) & 46(16,1,29)     & 39(7,0,32) & {\bf Total} \\\hline
\end{tabular}
\end{center}
(1) Miller et al. (1989); (2) Carini et al. (1990); (3) Carini et al. (1991); (4) Carini et al. (1992); 
(5) Carini \& Miller (1992); (6) Ghosh et al. (2001); (7) Romero et al. (2002); (8) Sagar et al. (2004); 
(9) Stalin et al. (2005) 
\end{table*}

Study of optical micro-variability of radio-loud AGNs (blazars) was done by 
several groups. The pioneer work in blazars optical intra-night variability is by 
Miller et al. (1989), Carini (1990), Carini et al. (1990, 1991, 1992) and Carini \& Miller 
(1992). First clear evidence of optical intra-night variability in BL Lacertae was reported by 
Miller et al. (1989). Carini et al. (1990) have observed  blazar OQ 530 in 4 nights (April 1-4,
1988), the source has shown micro-variability in all the four nights.  Carini et al. (1991) 
observed the blazar AP Librae in nine nights during March - May 1989. In three nights 
intra-night variability were seen. Carini \& Miller (1992) observed the blazar PKS 2155-304 
for continuous 4 nights (Sept. 25-28, 1988), and the micro-variability is seen on all these 
four nights. Carini et al. (1992) observed blazars OJ 287 and BL Lacertae for eighteen and 
fourteen nights respectively during (Nov. 1986 - March 1989). Out of 32 nights observations 
micro-variability is reported for 18 nights. Carini (1990), based on the study of a sample 
of 20 blazars, reported that the probability of seeing a significant micro-variability exceeds 
80\% if a source is monitored continuously for more than 8 hours.

\begin{figure}[h]
\hspace*{-0.2in}
\includegraphics[bb=18 144 592 718,width=4.0in,height=4.0in,clip]{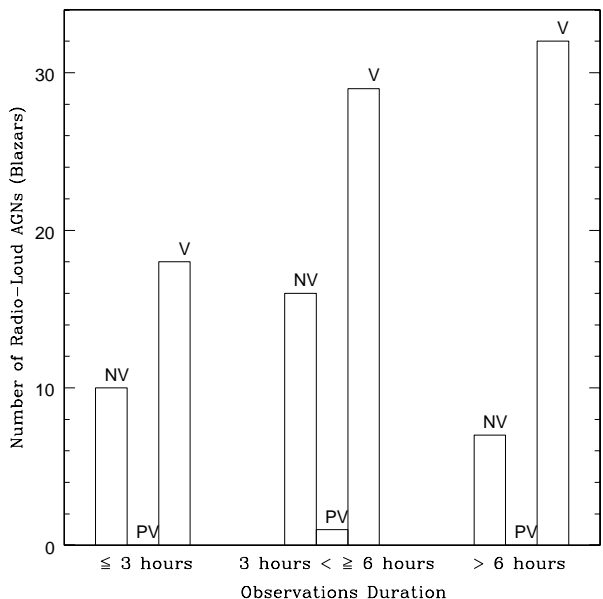}
\caption[]{Histogram of observing runs on radio-loud AGNs(blazars).
NV, PV and V denote the numbers of events the sources were detected: 
non variable, possible variable and variable respectively.}
\label{fig}
\end{figure}

Heidt \& Wagner (1996) studied optical intra-night variability in a sample of
34  radio-selected BL Lac objects from 1 Jy catalog. Observations were carried
out  during June 1990 to September 1993; each blazar was observed in seven
continuous  nights, 3 times in each night with time interval of 2 hours. In 28
out of 34 BL Lac  objects (82\%) intra-night variability was detected and 75\%
of the variable BL Lacs  changed significantly in time scale $<$ 6 hours. As
this data lacks continuity of lengthy data trains, we have not considered
these results in the statistical analysis in the present paper. 

About 140 intra-night light curves of a large sample of blazars were generated
in series of papers by Chinese group (ref. Bai et al. 1998, 1999; Dai et
al. 2001; Xie et al. 1999, 2001, 2002, 2004; Fan et al. 2001, 2004; Qian \& Tao
2004). Observations were made in more than two visual pass-bands and in a 
particular night two or more blazars were observed.  All the blazars in the
sample have shown micro-variations in several nights. The data in these
papers also lacks continuity of lengthy trains and hence not considered for 
further discussion and the statistical analysis in the present paper. 

Ghosh et al.(2001) have made observations on five blazars in seven nights
during November 05, 1997 $-$ December 29, 1998. Micro-variations were seen 
in four blazars. 

Romero et al.(2002) have reported observations of 4 EGRET blazars in
24 nights during the period 1997 to 2000. Intra-night variations were
reported for 18 nights in 3 blazars. One blazar was observed in 2 nights but
has not shown any intra-night variations.

Sagar et al. (2004) and Stalin et al. (2005) have recently done an extensive
search for intra-night optical variability in blazars. They have observed
nine BL Lac objects in 35 nights. All the sources have shown intra-night 
variations at least in one night observations. Out of 35 nights observations, 
intra-night variations  are seen in 20 nights.

We compiled the data from the literature on micro-variability of blazars 
(based on monitoring of radio-loud AGNs (blazars)) to study the statistics.
The data are listed in Table 5 and presented in the form of histogram in Fig. 12.
The data indicate that the events of occurrence of micro-variation in blazars
in time scale of less than 6 hour are $\sim$ 60-65 \%. If the blazar is observed
for more than 6 hour then the possibility of intra-night variability detection
is about 80-85 \%.

\section {Conclusions}

The new observations of RQQSOs reported here indicate clear evidence of the existence
of optical intra-night variability in the luminous RQQSOs. The compiled data of
all classes of AGNs, divided in three subgroups, show the presence of
intra-night variability in all the subclasses of AGNs. 

The popular model to explain micro-variations is shock-in-jet model (e.g.
Blandford \& Konigl 1979;  Scheuer \& Readhead 1979; Marsher 1980; Hughes et al.
1985; Marsher 1992; Marscher \& Gear 1985; Valtaoja et al. 1988; and Qian et al. 
1991). An important signature of the relativistic particle jets ejected by black 
holes is that their light is seen to fluctuate even on the time scale of less than 
an hour. This model is rather well accepted to explain micro-variability in radio-loud 
AGNs. The clear evidence of micro-variations in RQQSOs reported in the paper can be 
explained in all likelihood, relativistic particle jets are even ejected by the central 
engine of RQQSOs. However, probably most jets are quenched at the incipient stage 
itself, due to severe inverse-Compton losses inflicted by the intense photon field in 
the vicinity of the black hole. Thus, it appears to be no fundamental difference in the 
central engines of radio-quiet and radio-loud AGNs. 

Micro-variability reported here in the RQQSOs can also be supported by an alternative
standard model having numerous flares or hot spots on the accretion disk surrounding 
the central engine which can produce the micro-variations in quasars (e.g. 
Wiita et al. 1991, 1992; Chakrabarti \& Wiita 1993; Mangalam \& Wiita 1993). 

From the compiled catalog of micro-variations studies of radio-quiet and
radio-loud AGNs, we find that both the classes of AGNs have shown
micro-variations. Frequency of occurrence of micro-variations is least in
radio-quiet AGNs, highest in blazars and radio-loud AGNs
(excluding blazars) fall between these two extreme classes. Radio-quiet
AGNs exhibit micro-variations with maximum amplitude of about 10\% or less
whereas radio-loud AGNs (excluding blazars) show micro-variations with
amplitude of variation reaching to 50\% of the normal flux level 
with the frequency of occurrence being more than radio-quiet AGNs. On the 
other hand blazars show the extreme micro-variations with maximum amplitude 
of variation reaching  to $\sim$ 100\% of the normal flux level. Genearlly 
$\approx$ 10 \% and 35-40 \% radio-quiet AGNs and radio-loud AGNs (non-blazars) have
shown intra-night variations respectively. Any blazar, if observed continuously for 
less than 6 hours and more than 6 hours, the chances of seeing micro-variations are 
$\approx$ 60-65 \% and 80-85 \% respectively.

These results indicate that the energy generation mechanism and the environment 
around the central engine in different classes of AGNs may be similar, if not identical.
The standard models which explain the micro-variability in radio-loud AGNs viz. 
shock-in-jet models and accretion disk based models can also explain the micro-variability
behavior of RQQSOs.  
 
\begin{acknowledgements}

We thank the anonymous referee for his/her constructive critical comments
that helped to improve this paper. We are thankful to Profs. J. H. Fan and 
J. S. Bagla for reading the manuscript and making useful suggestions. The 
research work at the Physical Research Laboratory is funded 
by the Department of Space, Government of India. Department of Atomic Energy,
Government of India supported the research work at the Harish-Chandra Research 
Institute and at the Tata Institute of Fundamental Research.
IRAF is distributed by NOAO, USA. 

\end{acknowledgements}

\end{document}